\documentclass[a4paper, 10pt]{article}
\pdfoutput=1 
\usepackage[dvipsnames,svgnames,table]{xcolor} %
\usepackage{jcapmod}
\usepackage{verbatim}
\usepackage[T1]{fontenc} 
\usepackage[utf8]{inputenc}   

\definecolor{lightgray}{gray}{0.9}
\definecolor{Amber}{rgb}{1.0, 0.75, 0.0}
\definecolor{blizzardblue}{rgb}{0.67, 0.9, 0.93}
\definecolor{burgundy}{rgb}{0.5, 0.0, 0.13}
\usepackage{float}
\usepackage{graphicx}
\usepackage{hyperref}
\usepackage{amsmath}
\usepackage{xcolor}
\usepackage{tcolorbox}
\usepackage{amssymb}
\usepackage{physics}
\usepackage{microtype}
\usepackage[shortlabels]{enumitem}
\usepackage{cancel}
\usepackage{amsbsy}
\usepackage{color}
\usepackage[capitalise]{cleveref}
\usepackage{breakurl}
\usepackage{mathtools}
\usepackage{comment}
\usepackage{csquotes}
\usepackage{lmodern}
\usepackage{bm}
\usepackage{dsfont}
\usepackage{bbold}
\usepackage{empheq}
\usepackage{soul}
\usepackage{caption}
\usepackage{subcaption}
\usepackage[normalem]{ulem}

\pdfpageheight\paperheight
\pdfpagewidth\paperwidth

\usepackage[textwidth=0.71\paperwidth, textheight=0.8\paperheight]{geometry}

\renewcommand*{\vec}[1]{\bm{#1}}

\newcommand*{\mat}[1]{\bm{\mathsf{#1}}}

\let\phi\varphi

\DeclareMathOperator{\Var}{Var}
\newcommand{\identity}{\ensuremath{\mathds{1}}}

\newcommand{\LSS}{L_{\mathrm{LSS}}}

\newcommand*{\E}[1]{\texorpdfstring{\ensuremath{E_{#1}}}{E#1}}

\newcommand{\DKL}{D_{\mathrm{KL}}} 
\newcommand{\KL}[2]{\DKL\!\left(#1\,\|\,#2\right)}
\newcommand{\LR}{\Delta}     
\newcommand{\LCDM}{\Lambda\mathrm{CDM}}

\newcommand{\CP}{\mat{C}_{P}}
\newcommand{\CQ}{\mat{C}_{Q}}
\newcommand{\muP}{\vec{\mu}_P}
\newcommand{\muQ}{\vec{\mu}_Q}
\newcommand{\dmu}{\vec{\delta}} 
\newcommand{\Cc}{\mat{C}}

\renewcommand{\Tr}{\operatorname{tr}}
\newcommand{\TPR}{\mathrm{TPR}}
\newcommand{\FPR}{\mathrm{FPR}}

\usepackage[table]{xcolor}

\definecolor{burningsand}{RGB}{220, 148, 129}

\allowdisplaybreaks

\makeatletter
\DeclareRobustCommand{\rcite}[1]{%
  \rcite@aux#1,\@nil{#1}%
}
\def\rcite@aux#1,#2\@nil#3{%
  \if\relax#2\relax
    Ref.~\cite{#3}%
  \else
    Refs.~\cite{#3}%
  \fi
}
\makeatother

\tcbuselibrary{breakable}
\newtcolorbox{experiment}[1][Experiment protocol]{%
  breakable, colback=blizzardblue!18, colframe=dullblue,
  coltitle=white, fonttitle=\bfseries\small, title={#1},
  boxrule=0.6pt, left=2.2mm, right=2.2mm, top=1.2mm, bottom=1.2mm,
  before skip=7pt, after skip=7pt}
\renewcommand{\Tr}{\operatorname{tr}}

\subheader{IFT-UAM/CSIC-26-150}

\title{Beyond the Kullback--Leibler divergence: the likelihood-ratio distribution as a detectability measure}

\author[a]{Amirhossein Samandar,\textsuperscript{\(\dagger\)}}

\author[b,c]{Mikel Martin Barandiaran,}
\author[d]{Linn Htat Lu,}
\author[d]{Andrew~H.~Jaffe,}
\author[a,e]{Deyan P. Mihaylov,}
\author[a]{Glenn D. Starkman,}
\author[a]{Cynthia Trendafilova,}
\author[b,a]{Yashar Akrami,}
\author[f,g,h]{Stefano Anselmi,}
\author[b]{Javier Carr\'on Duque,}
\author[a]{Craig J. Copi,}
\author[i,a]{Fernando Cornet-Gomez,}
\author[j]{Arthur Kosowsky,}
\author[a]{Anna Negro,}
\author[k]{Ricardo~G.~Rodrigues,}
\author[l,a]{Andrius Tamosiunas}
\collaboration{(COMPACT Collaboration)}
\collaborationImg{\includegraphics[width = 0.1\textwidth]{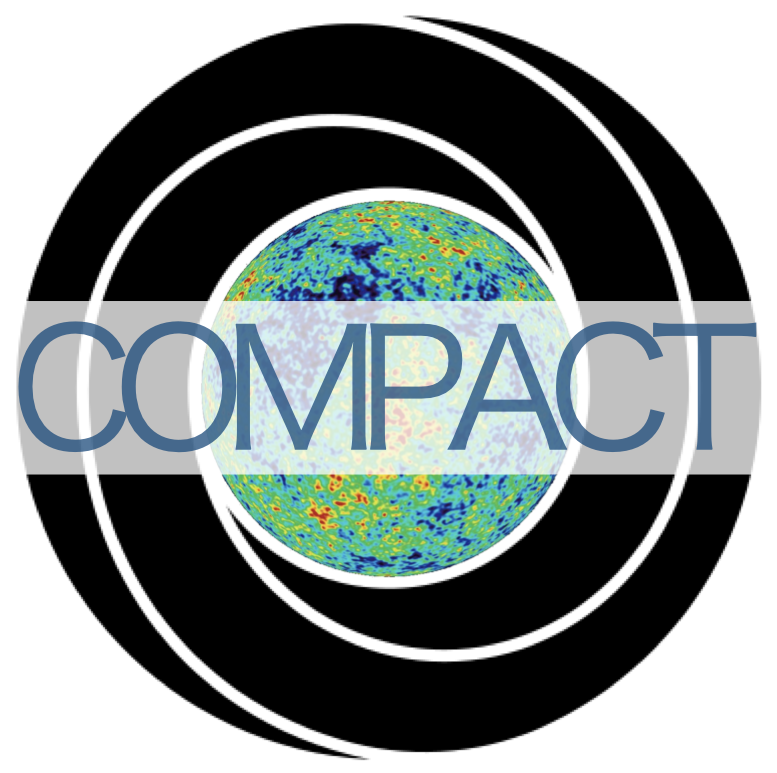}}

\affiliation[a]{CERCA/ISO, Department of Physics, Case Western Reserve University, 10900 Euclid Avenue, Cleveland, OH 44106, USA}
\affiliation[b]{Instituto de F\'isica Te\'orica (IFT) UAM-CSIC, C/ Nicol\'as Cabrera 13-15, Campus de Cantoblanco UAM, 28049 Madrid, Spain}
\affiliation[c]{Departamento de F\'isica Te\'orica, Universidad Aut\'onoma de Madrid, 28049 Madrid, Spain}
\affiliation[d]{Astrophysics Group \& Imperial Centre for Inference and Cosmology, Department of Physics, Imperial College London, Blackett Laboratory, Prince Consort Road, London SW7 2AZ, United Kingdom}
\affiliation[e]{Department of Astronomy, Faculty of Physics, Sofia University ``St Kliment Ohridski'', 5 James Bourchier Blvd, 1164 Sofia, Bulgaria}

\affiliation[f]{INFN, Sezione di Padova, via Marzolo 8, I-35131 Padova, Italy}
\affiliation[g]{Dipartimento di Fisica e Astronomia ``G. Galilei'', Universit\`a degli Studi di Padova, via Marzolo 8, I-35131 Padova, Italy}
\affiliation[h]{Laboratoire Univers et Th\'eories, Observatoire de Paris, Universit\'e PSL, Universit\'e Paris Cit\'e, CNRS, F-92190 Meudon, France}

\affiliation[i]{Departamento de F\'isica, Universidad de C\'ordoba, Campus Universitario de Rabanales, Ctra. N-IV Km. 396, E-14071 C\'ordoba, Spain}

\affiliation[j]{Department of Physics and Astronomy, University of Pittsburgh, Pittsburgh, PA 15260, USA}
\affiliation[k]{Departamento de F\'{i}sica, Universidade Estadual de Londrina, Rod. Celso Garcia Cid, Km 380, 86057-970, Londrina, Paran\'{a}, Brazil}
\affiliation[l]{Institute of Theoretical Astrophysics, University of Oslo, P.O. Box 1029 Blindern, N-0315 Oslo, Norway}

\emailAdd{amirhossein.samandar@case.edu}

\date{\today}

\abstract{
The Kullback--Leibler (KL) divergence is the expectation value of the log-likelihood ratio between two specified models, evaluated with respect to one of the two models. It provides an average measure of the distinguishability of the models. We examine its connection to discovery potential and the accuracy of parameter estimates when one model represents a null hypothesis, and the alternative is a family of models characterized by one or more continuous parameters. For data drawn from a multivariate Gaussian distribution, we derive the log-likelihood-ratio distribution and its cumulants under two such hypotheses. We then search over the alternative model's parameter space and use peak prominence, the height of a likelihood-profile peak above its surrounding baseline, as an operational detection statistic. Repeating this search on simulated data drawn from the null model allows us to calibrate the prominence threshold for a specified false-positive rate. We study a toy Gaussian model as well as three cosmological examples involving cosmic topology, the tensor-to-scalar ratio, and dark energy. In all these cases, detectability of a particular model depends not only on KL divergence, but also on the size of the model parameter space, the likelihood-ratio distributions, and the distance of the model from any parameter space boundaries. The relationship between KL divergence, discovery potential, and parameter estimation must therefore be calibrated for each family of models and detection statistic under consideration.
}


\begin{document}
\begingroup
\renewcommand{\thefootnote}{\fnsymbol{footnote}}
\footnotetext[2]{Corresponding author.}
\endgroup
\maketitle 
\flushbottom

\section{Introduction}
\label{sec:intro}

The Kullback--Leibler (KL) divergence quantifies the difference between two specified probability distributions \cite{Kullback:1951zyt,Kullback:1959}. For a hypothesized probability density $P$ and a reference probability density $Q$ over data $\vec a$,
\begin{equation}
\KL{P}{Q}=\int \mathrm{d}\vec a\,P(\vec a)
\ln\frac{P(\vec a)}{Q(\vec a)},
\label{eq:Kl_gen_definition}
\end{equation}
which is also referred to as the relative entropy of $P$ with respect to $Q$. 
By definition, the KL divergence is non-negative and vanishes if and only if the two densities are identical almost everywhere. It quantifies the expected information gain, measured in nats, of using $P$ rather than $Q$ when the data are drawn from $P$.

The KL divergence appears in several related contexts. In hypothesis testing, it characterizes how rapidly the probability of missed detection decreases as independent data accumulate \cite{Cover:2005lom}. With $Q$ as the null and $P$ as the alternative, this rate is set by $\KL{P}{Q}$, emphasizing that the ordering of the distributions matters. In Bayesian inference, the divergence between a posterior and its prior quantifies the information supplied by a dataset and enters the Bayesian complexity used to assess the effective number of constrained parameters \cite{Kunz:2006mc,Trotta:2008qt}. Relative entropy also underlies tests of consistency between cosmological datasets \cite{Nicola:2018rcd}, including the \emph{Surprise}, which measures the difference between the divergence obtained for a particular dataset and its expectation over possible datasets \cite{Seehars:2014ora,Grandis:2015qaa,Mello:2024tor}. These applications raise a natural question for model comparison: is there a given value of the KL divergence that corresponds to a practically detectable difference between two competing models?

The literature uses different rules of thumb regarding the connection between KL divergence and detectability. In cosmic topology, \rcite{Fabre:2013wia} applies the Jeffreys scale \cite{Jeffreys:1939xee,Trotta:2008qt} to the expected log Bayes factor, which equals $D_{\rm KL}$ for the fixed models under consideration, and adopts $D_{\rm KL}=1$ as the threshold of detectability. This criterion has since been used in forecasts by the COMPACT collaboration \cite{COMPACT:2022gbl,COMPACT:2024cud,COMPACT:2025tqc,COMPACT:2024dqe,COMPACT:2026pqs,COMPACT:2026nez,COMPACT:2026uzi}, with \rcite{COMPACT:2022gbl} further associating $D_{\rm KL}\gtrsim 3$ with a strong detection by appeal to Wilks' theorem,  which calibrates the likelihood ratio for use in frequentist hypothesis testing. On the other hand,~\rcite{Kunz:2005wh} relates the KL divergence to the expected log-likelihood ratio and reports its numerical values using base-10 logarithms. They argue that distinguishing two topologies will be very difficult when this value is less than $2$, corresponding to $\DKL\leq2\ln10\simeq4.6$ nats.

Meanwhile, in dark matter direct detection, \rcite{Herrera:2024zrk,Herrera:2026lqi} use the KL divergence to bound departures of the dark matter velocity distribution from a Maxwell--Boltzmann reference, considering values such as $D_{\rm KL}=0.1$ and $1$. Numerical criteria involving similar values also appear in comparisons of posterior distributions, although their purposes differ. In \rcite{Alfradique:2025vef}, $0.1$ and $1$ mark the boundaries between negligible, moderate, and substantial posterior divergence, whereas \rcite{Pellejero-Ibanez:2019enw,Rocher:2023zyh} use $D_{\rm KL}<0.1$ to monitor convergence of successive posterior approximations in likelihood emulation. In \rcite{Bevins:2025srk}, $D_{\rm KL}<1$ between the original and emulator-derived posteriors is instead a target for limiting incorrectly inferred information. In simulated galaxy surveys, \rcite{Oayda:2024voo} finds that the same posterior-to-prior divergence can yield different significances for a specified dipole-amplitude constraint as the survey geometry changes. These criteria concern different distributions and different inferential questions; consequently, their numerical values therefore do not, by themselves, provide detection thresholds that can be transferred from one analysis to another.

Nothing in \cref{eq:Kl_gen_definition} privileges a boundary of $D_{\rm KL}=1$, and a given value of the KL divergence does not generally correspond to a fixed probability of correctly favouring $P$ over $Q$. The reason is that $\KL{P}{Q}$ is an expectation of the log-likelihood ratio over data drawn from $P$. 
However, for a fixed experimental design, each repetition produces one realization of the data, and hence one realized log-likelihood ratio. The same reasoning applies to fields such as cosmology, where, within the standard description, only a single realization is observable. A large divergence implies that $P$ is favoured strongly on average over repeated realizations, but does not by itself specify how frequently an individual realization might instead favour $Q$.

Detectability therefore depends on the distribution of the log-likelihood ratio under both hypotheses. Its width and shape determine how frequently a realization exceeds a chosen decision threshold, giving the detection probability under $P$ and the false-positive rate under $Q$. For a fixed pair of models, the Neyman--Pearson lemma identifies the likelihood-ratio test as the most powerful test at a specified false-positive rate~\cite{Neyman:1933wgr}; the KL divergence supplies its mean under $P$, rather than its full distribution. The importance of calibrating a model-comparison statistic through its sampling distribution has also been emphasized by \rcite{Amendola:2024prl}, which derives sampling distributions for Bayes factors and uses them to construct frequentist tests. Here we focus on the log-likelihood ratio for specified models and on how its interpretation changes when the alternative is searched over a parameter family.

When the alternative is a family $P(\theta)$, a search evaluates the likelihood over the unknown parameter or parameters $\theta$. The question then extends beyond comparing two specified models to determining whether the data favour any member of the alternative family and how accurately its parameters can be recovered. A KL divergence computed for one fixed comparison does not directly provide the estimated error or parameter-recovery accuracy of this search. Even data drawn from the null can produce a preferred location in the scanned domain, because the same random realization is compared with multiple, correlated alternatives. Consequently, three questions must be distinguished: whether any member of the family is favoured sufficiently strongly over the null, where the preferred parameters lie, and how reliably those parameters recover the generating model. A detected peak need not imply accurate parameter recovery. Moreover, maximizing a likelihood over a family is distinct from computing its Bayesian evidence, which requires integration over a specified prior, and in general does not exist for improper (e.g., infinitely wide uniform) priors. The possibility that the null lies on a boundary of the parameter space, together with the look-elsewhere effect, must therefore be considered when calibrating the search statistic.

We address a general problem in statistical inference: how to translate the KL divergence between the probability densities of two models into a forecast of discovery potential and parameter recovery when the alternative is searched over a continuous parameter space. Building on established statistical tools~\cite{Self:1987,Protassov:2002sz,Gross:2010qma}, we develop a prescription that connects KL divergence and log-likelihood-ratio distributions to the detection probabilities and inferred parameters of a specified search. Simple cosmological case studies make this connection concrete, illustrating how the model family, parameter boundaries, and detection criterion enter the interpretation of a KL-based forecast.

In this paper, we connect fixed-model distinguishability to the detection and parameter-recovery properties of a search. For multivariate Gaussian data, we characterize the log-likelihood-ratio distribution under both hypotheses. We then examine one-parameter searches using peak prominence, the height of a likelihood-profile peak above its surrounding baseline, as an operational detection statistic. In future work, we will explore its extension to multi-parameter searches and the calibration of the corresponding detection thresholds. Simulations under the null calibrate its false-positive rate, while simulations under the alternatives measure detection probabilities and the distribution of accepted peak locations.
We apply this procedure to four experiments: a toy Gaussian covariance model, the search for cosmic topology with a cubic three-torus \E{1} compared with its infinite covering space \E{18}, the tensor-to-scalar ratio inferred from CMB $B$ modes, and a dark energy equation of state parameter inferred from supernova distance moduli. 

The paper is organized as follows. \Cref{sec:kl} introduces the log-likelihood ratio, its relation to the KL divergence, and the Gaussian mean- and covariance-shift cases. \Cref{sec:framework} derives the corresponding distributions and cumulants under both hypotheses, develops the parameter-search procedure, discusses Wilks' theorem and the look-elsewhere effect, and defines peak prominence. \Cref{sec:experiments} presents the four experiments, calibrates their false-positive rates, and compares detection probabilities and accepted peak locations.

\section{The KL divergence and log-likelihood ratio}
\label{sec:kl}

Consider two fully specified models for the data $\vec a$, with probability densities $P(\vec a)$ and $Q(\vec a)$. Taking $Q$ as the null hypothesis and $P$ as the alternative, the Neyman--Pearson lemma \cite{Neyman:1933wgr} states that the likelihood-ratio test is the most powerful test at a given significance level for this pair of hypotheses. We define the log-likelihood ratio (\emph{LLR}) as
\begin{equation}
\Delta(\vec a)\equiv\ln\frac{P(\vec a)}{Q(\vec a)},
\label{eq:LLRdef}
\end{equation}
Positive values favour $P$ over $Q$, while negative values favour $Q$. For these fully specified models, $\LR$ is also the log Bayes factor; including the prior model probabilities $\pi_P$ and $\pi_Q$ gives the log posterior odds, $\LR(\vec a)+\ln(\pi_P/\pi_Q)$. 

Taking the expectation of \cref{eq:LLRdef} and using \cref{eq:Kl_gen_definition}, we obtain
\begin{equation}
\mathbb{E}_{P}[\LR]=\KL{P}{Q},
\qquad
\mathbb{E}_{Q}[\LR]=-\KL{Q}{P}.
\label{eq:kl_means}
\end{equation}
Thus, $\KL{P}{Q}$ is the expected value of the \emph{LLR} when the data are drawn from $P$, while $-\KL{Q}{P}$ is the expectation of the same statistic when the data are drawn from $Q$. This relationship holds independently of the form of the two distributions.

\subsection{Definition and Gaussian form}
\label{sec:kl-def}

\Cref{eq:Kl_gen_definition} makes no reference to any particular family of distributions, and neither does the interpretation of $\KL{P}{Q}$ as an expected information gain. Since Gaussian models are widely used in scientific data analysis, we give the closed-form expression for this family.
Consider two multivariate Gaussian models for the real-valued\footnote{Restricting to real-valued data entails no loss of generality for the cases of interest here. 
Even when a complex representation of the data is used, such as the spherical-harmonic coefficients $a_{\ell m}$ of the CMB, the expressions in this paper then apply directly to the complex data vectors, provided they contain all $-\ell\le m\le\ell$ and transposes are replaced by Hermitian conjugates.}
 data vector $\vec{a}\in\mathbb{R}^{N}$,
\begin{equation}
Q:\ \vec{a}\sim\mathcal{N}(\muQ,\CQ),
\qquad
P:\ \vec{a}\sim\mathcal{N}(\muP,\CP),
\qquad
\dmu\equiv\muP-\muQ\,,
\label{eq:twohyp}
\end{equation}
where $\mathcal{N}(\vec{\mu},\mat{C})$ is a multivariate normal distribution with mean $\vec{\mu}$ and covariance matrix $\mat{C}$. We assume that both covariance matrices are positive definite, and $N$ is the dimension of the data vector. 

Substituting the Gaussian densities in \cref{eq:twohyp} into \cref{eq:LLRdef} gives
\begin{equation}
\boxed{\;
\LR(\vec{a})
=\tfrac12(\vec{a}-\muQ)^{\top}\CQ^{-1}(\vec{a}-\muQ)
-\tfrac12(\vec{a}-\muP)^{\top}\CP^{-1}(\vec{a}-\muP)
+\tfrac12\ln\frac{\det\CQ}{\det\CP}.\;}
\label{eq:LLR}
\end{equation}
The \emph{LLR} is therefore an inhomogeneous quadratic form in the
data. Writing $\vec x=\vec a-\muP$, its quadratic term is
$\tfrac12\vec x^\top(\CQ^{-1}-\CP^{-1})\vec x$,
while its linear term is $\dmu^\top\CQ^{-1}\vec x$;
the remaining terms are constant. In this centred form,
the covariance difference determines the quadratic contribution,
and the mean difference introduces the linear contribution.

Taking $P$ as the underlying distribution of a given data and $Q$ as its approximation, the KL divergence in \cref{eq:Kl_gen_definition} becomes
\begin{equation}
\begin{aligned}
\KL{P}{Q}
&=\tfrac12\!\left[
\Tr(\CQ^{-1}\CP)-N
+\ln\frac{\det\CQ}{\det\CP}
+\dmu^\top\CQ^{-1}\dmu
\right]\\
&=\underbrace{\tfrac12\sum_{i=1}^{N}
(\lambda_i-1-\ln\lambda_i)}_{\text{covariance}}
+\underbrace{\tfrac12\dmu^\top\CQ^{-1}\dmu}_{\text{mean}},
\end{aligned}
\label{eq:KLgauss}
\end{equation}
where $\lambda_i>0$ are the eigenvalues of $\CQ^{-1}\CP$. In the opposite ordering,
\begin{equation}
\KL{Q}{P}
=\tfrac12\sum_{i=1}^{N}
\big(\lambda_i^{-1}-1+\ln\lambda_i\big)
+\tfrac 12\dmu^\top\CP^{-1}\dmu .
\label{eq:KLqp}
\end{equation}

Both contributions to \cref{eq:KLgauss} are non-negative. Each term $\lambda_i-1-\ln\lambda_i$ vanishes only at $\lambda_i=1$, while the positive-definite quadratic form $\dmu^\top\CQ^{-1}\dmu$ vanishes only at $\dmu=\vec{0}$. Thus, $\KL{P}{Q}=0$ if and only if $\CP=\CQ$ and $\muP=\muQ$. The same reasoning applies to $\KL{Q}{P}$. There are two cases in which only one of these contributions remains, as follows.

\paragraph{Pure covariance shift ($\dmu=\vec{0}$):}
When the two distributions have the same mean, the mean contribution vanishes and \cref{eq:KLgauss} reduces to the covariance-only expression.
Any difference between the models then lies in their variances and correlations. The common mean need not be zero and can be subtracted without changing the divergence. With $\vec x=\vec a-\muP$, \cref{eq:LLR} becomes
\begin{equation}
\LR(\vec a)
=\tfrac12\vec x^\top(\CQ^{-1}-\CP^{-1})\vec x
+\tfrac12\ln\frac{\det\CQ}{\det\CP},
\label{eq:lambda-same_mean}
\end{equation}
so the dependence on the centred data is entirely quadratic.

\paragraph{Pure mean shift ($\CP=\CQ\equiv\Cc$):}
If instead the two distributions have the same covariance, then all $\lambda_i=1$ and the covariance term drops out. In this case,
\begin{equation}
\KL{P}{Q}=\KL{Q}{P}
=\frac{1}{2} \dmu^\top \Cc^{-1} \dmu,
\label{eq:KLmeanshift}
\end{equation}
so the divergence depends only on the difference between the means, weighted by the inverse of their common covariance, and is symmetric under exchanging $P$ and $Q$. For two one-dimensional Gaussians with common variance $\sigma^2$, this reduces to $D_{\rm KL}=\Delta\mu^2/(2\sigma^2)$. Thus, $D_{\rm KL}=1$ corresponds to a mean separation of $\sqrt{2}\sigma\simeq1.4\sigma$.
Moreover, the quadratic terms in \cref{eq:LLR} cancel, leaving
\begin{equation}
\LR(\vec a)
=\dmu^\top\Cc^{-1}(\vec a-\muQ)
-\tfrac12\dmu^\top\Cc^{-1}\dmu.
\label{eq:LLRmeanshift}
\end{equation}
The data-dependent term is the matched filter for the mean
difference $\dmu$, with weighting set by $\Cc^{-1}$.
The \emph{LLR} is therefore linear in the data up to a constant
offset. This distinction between quadratic and linear
dependence determines the form of the \emph{LLR} distribution,
which we derive under both hypotheses in \cref{sec:framework}.

\section{A distribution-level detectability framework}
\label{sec:framework}

As shown in \cref{eq:kl_means}, the KL divergence gives the mean of $\LR(\vec a)$ under $P$. However, knowing this mean does not generally tell us how often a realization will favour one model over the other, since those probabilities depend on the width and shape of the $\LR(\vec a)$ distribution. To connect the divergence to detection rates, we therefore need to characterise the distribution of $\LR(\vec a)$ under both hypotheses.
For the Gaussian pair considered here, this distribution can be obtained exactly by expressing \cref{eq:LLR} as a quadratic form with a linear term in independent Gaussian variables.
Here we apply standard results of generalized $\chi^2$ distributions \cite{Imhof1961, mathai1992} to the \emph{LLR} and examine how its distribution changes when the models differ only in their covariances or only in their means.

\subsection{The distribution of the log-likelihood ratio}
\label{sec:llr-dist}

To apply these results to $\LR(\vec a)$, we first express the data in terms of independent standard Gaussian variables. Under $P$, we write
$\vec{a}=\muP+\CP^{1/2}\vec{u}$, where
$\vec{u}\sim\mathcal{N}(\vec{0},\identity)$ and $\CP^{1/2}$ is the symmetric positive-definite square root of $\CP$.
Substituting this expression into \cref{eq:LLR} gives the quadratic kernel
$\tfrac12\CP^{1/2}(\CQ^{-1} -\CP^{-1})\CP^{1/2}$.
An orthogonal matrix $\mat{O}$ diagonalises this kernel, and the corresponding rotation preserves the independence and standard normal distribution of the components of $\vec{u}$. Thus, $\LR(\vec a)$ takes the form of 
\begin{equation}
\LR(\vec a)=\sum_{i=1}^{k}\big(\beta_i u_i^2+w_i u_i\big)+b,
\label{eq:genchi2-def}
\end{equation}
with
\begin{equation}
\begin{aligned}
\beta_i&=\tfrac12(\lambda_i-1),
\qquad
\vec{w}=\mat{O}^{\top}\CP^{1/2}\CQ^{-1}\dmu,\\
b&=\tfrac12\dmu^{\top}\CQ^{-1}\dmu
-\tfrac12\sum_i\ln\lambda_i,
\end{aligned}
\label{eq:llr-P-params}
\end{equation}
where $\lambda_i$ are the eigenvalues of $\CQ^{-1}\CP$, ordered to match the columns of $\mat{O}$.

Under $Q$, the same construction begins with $\vec{a}=\muQ+\CQ^{1/2}\vec{u}$ and gives $\tilde\beta_i=\tfrac12(1-\lambda_i^{-1})$,
$\tilde{\vec{w}}=\tilde{\mat{O}}^{\top}\CQ^{1/2}\CP^{-1}\dmu$,
and
$\tilde b=-\tfrac12\dmu^\top\CP^{-1}\dmu
-\tfrac12\sum_i\ln\lambda_i$.
Here $\CQ^{1/2}$ is again the symmetric positive-definite square root, and the columns of $\tilde{\mat{O}}$ diagonalise the kernel
$\tfrac12\CQ^{1/2}(\CQ^{-1}-\CP^{-1})\CQ^{1/2}$, with their ordering chosen to match the $\tilde\beta_i$. 
The cumulants of the \emph{LLR} distributions under the two hypotheses are:

\paragraph{Under $P$:}
\begin{empheq}[box=\fbox]{align}
\kappa_1^P=\mathbb{E}_P[\LR]
&=\tfrac12\sum_i(\lambda_i-1 -\ln\lambda_i)
+\tfrac12\dmu^\top\CQ^{-1}\dmu =\KL{P}{Q}, 
\label{eq:k1P}\\
\Var_P[\LR]=\kappa_2^P
&=\tfrac12\sum_i(\lambda_i-1)^2+\sum_iw_i^2,
\label{eq:k2P}\\
\kappa_3^P
&=\sum_i(\lambda_i-1)^3
+3\sum_iw_i^2(\lambda_i-1),
\label{eq:k3P}\\
\kappa_n^P
&=\tfrac{(n-1)!}{2}\sum_i
\Big[(\lambda_i-1)^n
+n\,w_i^2(\lambda_i-1)^{n-2}\Big],
\qquad n\ge2.
\label{eq:knP}
\end{empheq}

\paragraph{Under $Q$:}
\begin{equation}
\begin{aligned}
\kappa_1^Q &=-\KL{Q}{P},\\
\kappa_n^Q &=\tfrac{(n-1)!}{2}\sum_i
\Big[(1-\lambda_i^{-1})^n +n\,\tilde w_i^{\,2}(1-\lambda_i^{-1})^{n-2}\Big], \qquad n\ge2.
\end{aligned}
\label{eq:cumQ}
\end{equation}

The first cumulant under $P$ reproduces \cref{eq:KLgauss}, while under $Q$ it gives $-\KL{Q}{P}$, as required by the general identity in \cref{eq:kl_means}. The separation between these two means is therefore the Jeffreys divergence, $\KL{P}{Q}+\KL{Q}{P}$. However, this separation must be considered alongside the fluctuations under each hypothesis, whose variances are given by the corresponding second cumulants. The higher cumulants provide further information about the shape that the KL divergence does not contain.
For example, the skewness under $P$,
$\mathrm{skew}_P=\kappa_3^P/(\kappa_2^P)^{3/2}$, describes the asymmetry of the \emph{LLR} distribution and can be evaluated directly from the same spectral coefficients once they have been computed. We checked the first four cumulants in
\cref{eq:k1P,eq:k2P,eq:k3P,eq:knP,eq:cumQ} against direct Monte Carlo sampling of \cref{eq:LLR} in \cref{fig:example-profiles}.

The two cases introduced in \cref{sec:kl-def} now lead to distinct forms of the \emph{LLR} distribution:

\paragraph{Covariance-shift limit ($\dmu=\vec{0}$):}
When the means coincide, both $\vec{w}$ and $\tilde{\vec{w}}$ vanish, so the fluctuations in $\LR(\vec a)$ come entirely from the quadratic terms. For $n\ge2$, the cumulants reduce to

\begin{equation}
\begin{aligned}
\kappa_n^P&=\tfrac{(n-1)!}{2}\sum_i(\lambda_i-1)^n,\\
\kappa_n^Q&=\tfrac{(n-1)!}{2}\sum_i(1-\lambda_i^{-1})^n.
\end{aligned}
\label{eq:covariance-cumulants}
\end{equation}

The \emph{LLR} therefore follows a generalized $\chi^2$ distribution. Although odd cumulants can vanish through cancellations between positive and negative weights, the fourth cumulant is strictly positive whenever $\CP\neq\CQ$, so the distribution is not
exactly Gaussian.

\paragraph{Mean-shift limit ($\CP=\CQ=\Cc$):}
When the covariances coincide, all $\lambda_i=1$ and the quadratic weights vanish under both hypotheses. The \emph{LLR} is then linear in the Gaussian data, so its distribution is exactly Gaussian.
In particular, \cref{eq:cumQ,eq:knP} gives non-zero values for only the first two cumulants. Since
$\sum_iw_i^2=\sum_i\tilde w_i^{\,2}
=\dmu^\top\Cc^{-1}\dmu=2\KL{P}{Q}$,
we obtain
\begin{equation}
\boxed{
\begin{aligned}
\LR\big|_P&\sim\mathcal{N}\big(\KL{P}{Q},\,2\KL{P}{Q}\big),\\
\LR\big|_Q&\sim\mathcal{N}\big(-\KL{P}{Q},\,2\KL{P}{Q}\big).
\end{aligned}
}
\label{eq:LLRgauss}
\end{equation}
These distributions are exact for any mean separation, with no large-sample or small-shift approximation. Because the KL divergence is symmetric in this limit, their means have equal magnitudes and opposite signs, while their common variance is twice $\KL{P}{Q}$. If the means also coincide, the models are identical and $\LR(\vec a)=0$ for every realization.

\subsection{From a fixed comparison to a profile over parameters}
\label{sec:profile}
So far we have studied the distribution of $\LR(\vec a)$ for a fixed pair of models. In practice, the alternative is usually a family of models $P(\theta)$ indexed by an unknown parameter, and the data must be searched over $\theta$ before any statement about $P$ can be made.\footnote{Even more realistically, the alternative model will depend on various parameters $\bm{\theta}=(\theta_1,\dots,\theta_n)$. For the purposes of this work we adopt the simplifying assumption that it only depends on a single parameter.} We therefore evaluate the \emph{LLR} profile along the scanned domain $\Theta_{\rm scan}$ and record its maximum. We emphasize that, although the likelihood also enters Bayesian analyses, maximization alone neither introduces a prior nor produces a Bayesian evidence. 

Let $\theta$ denote the searched parameter, $\muP(\theta)$ and $\CP(\theta)$ the corresponding alternative model mean and covariance, and let $\muQ$ and $\CQ$ describe the fixed null model. For a data vector $\vec a$, using the Gaussian \emph{LLR} in \cref{eq:LLR}, we define its maximum over the scanned parameter range as
\begin{equation}
T(\vec a)=\max_{\theta\in\Theta_{\rm scan}}\LR(\vec a;\theta).
\label{eq:peak}
\end{equation}
For every fixed $\theta$, the results of
\cref{sec:llr-dist} characterize the corresponding \emph{LLR} distribution, including its exactly Gaussian mean-shift limit.
They do not, however, determine the distribution of $T$. The trial \emph{LLR}s are computed from the same data and are therefore correlated, so the maximum of these \emph{LLR}s does not generally follow the same distribution as any individual trial \emph{LLR}. Determining the distribution of this maximum over a correlated random process is the central difficulty in searches where the signal parameter is present only under the alternative \cite{Davies:1977,Davies:1987zz,Gross:2010qma}.
 
\subsection{Wilks' theorem, boundary nulls, and the look-elsewhere effect}
\label{sec:lee}
The maximum in \cref{eq:peak} is closely related to the standard profile-likelihood construction \cite{Cowan:2010js}.
In a general analysis, nuisance parameters are maximized over at each trial value of the parameter of interest. In our experiments, they are instead fixed or eliminated before the scan, leaving a one-parameter likelihood. We define $\Delta\chi^2(\theta)$, the likelihood drop relative to the best fit, as
\begin{equation}
\Delta\chi^2(\theta)
\equiv -2\ln\frac{P(\vec a;\theta)}{\max_{\theta'\in\Theta_{\rm scan}}P(\vec a;\theta')} = 2[T(\vec a)-\LR(\vec a;\theta)].
\label{eq:plr}
\end{equation}
This likelihood drop therefore differs from the \emph{LLR} against $Q$. Neither is a Bayesian evidence, which requires integration over a specified prior.

For parameter estimation, Wilks' theorem connects this likelihood drop to an asymptotic reference distribution. For a regular model with one identifiable parameter, a true value $\theta_{\rm true}$ in the interior of the scanned domain, a smooth likelihood, and non-zero Fisher information, the sampling distribution of the likelihood-drop statistic evaluated at the true parameter approaches a chi-squared distribution with one degree of freedom \cite{Wilks:1938dza}:
\newcommand{\widesim}[2][1.5]{
  \mathrel{\overset{#2}{\scalebox{#1}[1]{$\sim$}}}
}
\begin{equation}
\Delta\chi^2(\theta)
\widesim[3]{\mathrm{asymp.}}\chi^2_1\,, 
\label{eq:wilks}
\end{equation}
where $\widesim[3]{\mathrm{asymp.}}$ denotes convergence in distribution in the large-sample limit, and $\chi^2_1$ denotes the chi-squared distribution with one degree of freedom. 
The sampling distribution describes the variation of the likelihood-drop statistic across repeated datasets generated at the same $\theta_{\rm true}$.
This result, \eqref{eq:wilks}, underlies the familiar likelihood-drop intervals, such as $\Delta\chi^2\leq1$ for an approximately $68.3\%$ confidence interval in the one-parameter regular limit. The same prescription is used to quote frequentist constraints from profile likelihoods in cosmology \cite{Planck:2013nga,Herold:2021ksg,Campeti:2022vom,Herold:2024enb}. 
This prescription provides approximate confidence intervals around the best-fit parameter when the true parameter lies in the interior of the domain and the regularity conditions of \cref{eq:wilks} hold.

For detection, we instead ask whether any allowed alternative is preferred sufficiently strongly over $Q$. This question differs from the earlier discussion of parameter estimation in two ways. First, the null lies on the boundary of every physical model family considered here, so when $Q$ is true, the true parameter is not an interior point of the domain. Even for a single, known signal location, $2\LR$ then follows a mixture of $\chi^2$ distributions rather than $\chi^2_1$ \cite{Chernoff:1954eli,Self:1987}. Second, the signal location $\theta$ is undefined when there is no signal, so it cannot be identified under the null \cite{Davies:1977,Davies:1987zz}. These two properties violate the interior-point and identifiability conditions of Wilks' theorem stated above \cite{Protassov:2002sz}. As a result, $2T(\vec a)$ is not asymptotically $\chi^2$-distributed under $Q$ \cite{Ranucci:2012ed,Algeri:2016gtj}, and its distribution must be found by other means.

To address these failures, one can account for the \emph{look-elsewhere effect} (LEE) \cite{Lyons:2008hdc,Gross:2010qma}. It is best seen by asking what a signal-free realization does under the same search. A realization of the null model still produces a fluctuating profile $\LR(\vec a;\theta)$ whose global maximum is typically positive, because scanning many values of $\theta$ gives the random structure of that one realization many chances to align with the model at some $\theta$. A candidate peak must therefore be judged not against zero but against the height that the best peak of a null realization typically reaches. In the standard treatment, the \emph{local} significance at fixed $\theta$ and the \emph{global} significance over the scan are related by a \emph{trials factor}, heuristically the number of independent chances the scan gives noise to produce a false peak; the wider the search, the larger the trials factor, and the taller a peak must be before it counts as a genuine detection \cite{Gross:2010qma}. Most closely related to our approach, self-calibration estimates the correction directly from the peaks in the same dataset. In its simplest form, the LEE-corrected $\chi^2$ of a peak is approximated by subtracting the $\chi^2$ of the highest noise-induced peak from the $\chi^2$ of the peak of interest \cite{Bayer:2021lhk}.

Our setting permits a simpler treatment. The profile is a function of a single parameter, no nuisance parameters remain, and the null model is fully specified, so the null distribution of any statistic built from the profile can be sampled directly from realizations of $Q$ over the same scanned range and grid. The trials-factor bookkeeping is then replaced by a direct Monte Carlo calibration against null realizations \cite{Hope:1968}; with $N_{\rm null}$ realizations, false-positive rates below about $1/N_{\rm null}$ cannot be resolved. To distinguish a genuine detection, we ask whether the profile presents a single dominant, isolated peak, measured by its \emph{prominence}, and read the associated false-positive rate directly from realizations of $Q$.

\subsection{The peak as the detection signature}
\label{sec:peak}

When $\theta$ is scanned, the profile $\LR(\vec{a};\theta)$ is expected to reach a maximum near $\theta=\theta_{\rm true}$. This peak is the operational detection signature and yields three separable diagnostics, which we advocate reporting jointly:
\begin{enumerate}[label=(\roman*),leftmargin=2em,itemsep=1pt,topsep=2pt]
    \item the \emph{peak position} $\hat\theta=\arg\max_\theta\LR(\vec a;\theta)$, locating the candidate signal in parameter space thereby answering the question: is $\theta_{\rm true}$ recovered?;
    \item the \emph{peak height} $T(\vec a)$, the maximized \emph{LLR} of the family $P(\theta)$ against the null $Q$; and
    \item the \emph{peak prominence}, the height of the peak above its surrounding profile (defined in \cref{eq:prominence}), which we compare with a calibrated threshold to declare a detection.
\end{enumerate}
 
Let $\hat\theta$ be the global maximum of $\LR(\vec a;\theta)$. Following the standard topographic definition~\citep{Kirmse:2017}, descend from $\hat\theta$ in each direction and let $\theta_{\rm col}$, the peak's \emph{key col}, be the higher of the two minima reached before the profile either exceeds $\LR(\vec a;\hat\theta)$ again or hits the boundary of the explored domain. The raw prominence is
\begin{equation}
\mathrm{prom}(\hat\theta)\;\equiv\;\LR(\vec a;\hat\theta)-\LR(\vec a;\theta_{\rm col}).
\label{eq:prominence}
\end{equation}
For a sampled profile, this is the definition implemented in \texttt{scipy.signal.peak\_prominences} \cite{Virtanen:2019joe}.
 
\subsection{Prominence as a peak statistic}
\label{sec:whyprominence}
The maximum \emph{LLR} measures the strongest preference for a trial model over $Q$, but it does not describe how that maximum stands above its surrounding profile. Peak prominence addresses this second question,
defining an operational detection signature by the rise of a local maximum above a baseline within the same realization. This choice allows us to study the occurrence of peaks and their locations separately from the value of the \emph{LLR} at the generating parameter. It is a choice of detection statistic, whose performance must be
established by calibration. The profile is not noise-free: even when no instrumental noise is added, the random realization of the data still produces fluctuations in the profile and can generate peaks under the null model $Q$.
 
The overall scale of $\LR(\vec a;\theta)$ varies from one realization to another, and realizations of $Q$ also yield $\LR>0$ at some $\theta$ with non-zero probability, so a positive maximum is not by itself evidence for a signal.
We declare a detection when the prominence of the highest peak, which is also the global maximum of $\LR$, whose height is $T(\vec a)$, exceeds a specified threshold,
\begin{equation}
\mathrm{prom}(\hat\theta)\;\ge\;\mathrm{prom}_{\rm thr},
\label{eq:detection-criterion}
\end{equation}
which separates a dominant maximum from a flat profile, in which no $\theta$ is preferred ($\mathrm{prom}\to 0$). The threshold $\mathrm{prom}_{\rm thr}$ is fixed by calibrating against profiles computed from realizations of the null model $Q$, over the same scanned range and grid, so that the false-detection rate over those realizations is held at the desired level. This is the direct, simulation-based counterpart of setting a $p$-value threshold in the standard pipeline \cite{Hope:1968,Algeri:2016gtj}. A successful \emph{Parameter recovery} is then declared if the detected peak location $\hat\theta$ lies within a prescribed tolerance of the true value $\theta_{\rm true}$.

\section{Case studies}
\label{sec:experiments}
\label{sec:experiment-overview}

In this section, we study four experiments that share the same statistical structure. For a given fixed null $Q$ and a one-parameter family $P(\theta)$, how does searching the profile $\LR(\vec a;\theta)$ change what can be inferred from a single realization? Each family approaches $Q$ at an edge of its parameter domain, and we use $D_{\mathrm{KL}}(\theta)\equiv\KL{P(\theta)}{Q}$ to place the alternatives on a common information scale. In each case the map $\theta\mapsto\KL{P(\theta)}{Q}$ is monotonic over the scanned range, so the scanned parameter can be exchanged for the divergence of the model that generated the data. This is what makes the experiments comparable.

The experiments differ in the shape of the distribution of $\LR$ at fixed $\theta$, as discussed in \cref{sec:framework}. Three of them place the signal in the covariance of zero-mean data, the covariance-shift limit of \cref{sec:llr-dist}, in which $\LR$ is a generalized $\chi^2$ with non-vanishing cumulants at every order. Within this covariance group, which comprises the controlled Gaussian model, the cubic three-torus \E{1}, and the tensor-to-scalar ratio $r$, the degree of non-Gaussianity is itself tunable. The fourth experiment places the signal in the mean at fixed covariance, the mean-shift limit, in which $\LR$ is exactly Gaussian (\cref{eq:LLRgauss}). The dark energy experiment using supernova data and $w_a=0$ plays a distinct role in this set. Because its fixed-parameter problem is exactly Gaussian, $\KL{P}{Q}$ fully determines the operating characteristic of a test between $Q$ and any single $P(w_0)$.

These are semi-realistic and deliberately simplified setups. Each is chosen so that $D_{\mathrm{KL}}(\theta)$, the cumulants of $\LR$, and the profile over $\theta$ can be computed in closed form or by direct linear algebra, and so that the statistical behavior of the search is visible. They are not new measurements of topology, $r$-ratio, or $w_0$, and no cosmological conclusion should be drawn from them. The structure of the four setups is summarized in \cref{tab:experiments}.

\begin{table}[t]
\centering
\footnotesize
\setlength{\tabcolsep}{4pt}
\renewcommand{\arraystretch}{1.3}
\begin{tabular}{p{2.1cm}p{2.9cm}p{2.9cm}p{3.2cm}p{2.9cm}}
\hline
& \raggedright Toy Gaussian (\cref{sec:toy-gaussian})
& \raggedright Cubic \E{1} (\cref{sec:topology})
& \raggedright Tensor-to-scalar ratio (\cref{sec:rratio})
& \raggedright Dark energy (\cref{sec:sn}) \tabularnewline
\hline
\raggedright Data
& \raggedright $a_{\ell m}$, $2\le\ell\le40$
& \raggedright $a^T_{\ell m}$, $2\le\ell\le40$
& \raggedright $a^B_{\ell m}$, $2\le\ell\le40$
& \raggedright distance moduli of $N$ SNe \tabularnewline
\raggedright Signal enters
& \raggedright covariance (diagonal)
& \raggedright covariance (non-diagonal)
& \raggedright covariance (diagonal)
& \raggedright mean (fixed covariance) \tabularnewline
\raggedright $\LR$ at fixed $\theta$
& \raggedright generalized $\chi^2$; $\lambda_\ell<1$, $\mathrm{skew}<0$, varies with $t$
& \raggedright generalized $\chi^2$; spectrum numerical
& \raggedright generalized $\chi^2$; $\lambda_\ell>1$, $\mathrm{skew}_P>0$, independent of $r$
& \raggedright exactly Gaussian \tabularnewline
\raggedright Scanned $\theta$
& \raggedright $t\in[2.6,25]$
& \raggedright $L\in(\LSS,1.4\,\LSS]$
& \raggedright $r\in[0,0.036]$
& \raggedright $w_0\in[-1,w_0^{\max}]$ \tabularnewline
\raggedright Null at
& \raggedright $t\to\infty$ (beyond grid)
& \raggedright $L\to\infty$ (beyond grid)
& \raggedright $r=0$
& \raggedright $w_0=-1$ \tabularnewline
\raggedright Far edge set by
& \raggedright choice of divergence range
& \raggedright circles in the sky
& \raggedright existing upper limit on $r$
& \raggedright end of acceleration, $q_0=0$ \tabularnewline
\hline
\end{tabular}
\caption{Structure of the four numerical experiments. ``Far edge'' is the edge
of the scanned range opposite the null, where $\KL{P(\theta)}{Q}$ is largest.}
\label{tab:experiments}
\end{table}

For each experiment, we generate ensembles under $Q$ and under selected $P(\theta_{\rm true})$, evaluate the full profile for every realization, and apply the same peak-selection procedure to both ensembles. We first use the null ensembles to establish the relation between the prominence threshold and the false-positive rate. We then compare the locations of accepted peaks and the number of detections under the alternatives. Finally, we use receiver operating characteristic (ROC) curves, which trace the true-positive rate against the false-positive rate as the threshold is varied \cite{Fawcett:2006}, to examine how the two rates change together.

In these simulations, we know which model generated each realization, denoted by $\theta_{\rm true}$. If a peak at $\hat\theta$ is accepted as a detection, we express both locations through their KL divergence from $Q$:
\begin{equation}
\DKL({\rm true})\equiv \KL{P(\theta_{\rm true})}{Q},\qquad
\DKL({\rm peak})\equiv \KL{P(\hat\theta)}{Q}.
\label{eq:peak-divergence}
\end{equation}
Comparing these quantities shows whether the accepted peaks recover the underlying model that generated the realization and whether they are more or less distinguishable from the null than the generating model.

\subsection{A toy Gaussian model}
\label{sec:toy-gaussian}
 
We begin with a phenomenological model whose covariance is diagonal in harmonic space. This provides a controlled setting for examining the behavior of the search before turning to the physical models of \cref{sec:topology,sec:rratio,sec:sn}. We use real zero-mean Gaussian coefficients $a_{\ell m}$ with $2\leq\ell\leq40$, $|m|\leq\ell$, and diagonal covariance depending only on the value of $\ell$, under both hypotheses. This is equivalent to the distribution of CMB temperature anisotropies with covariance $\langle\left|a_{\ell m}\right|^2\rangle=C_\ell$.
The null is $Q:\vec a\sim\mathcal N(\vec0,\CQ)$, while the alternative is $P(t):\vec a\sim\mathcal N(\vec0,\CP(t))$, with
\begin{equation}
C_\ell^Q=C_\ell,\qquad C_\ell^P(t)=\lambda_\ell(t)C_\ell,\qquad
\lambda_\ell(t)=1-A\exp\!\left[-\frac{\ell^2t^2}{2\ell_c^2}\right],\quad A=0.8,\quad \ell_c=22.
\label{eq:toy-lambda}
\end{equation}
The constant $\ell_c$ sets the multipole scale of the suppression, while $C_\ell$ denotes the reference angular power spectrum. At each $\ell$, all $g_\ell=2\ell+1$ independent real degrees of freedom have the same variance. For each generating model, we draw an ensemble of realizations and keep each realization fixed throughout its parameter scan, preserving the correlations between profile values at different trial parameters.

We scan $t\in[2.6,25]$, with the lower endpoint setting the largest KL divergence explored. The exact null is approached as $t\to\infty$ and therefore lies beyond the finite grid. The KL divergence and \emph{LLR} follow the covariance-shift expressions in \cref{sec:llr-dist,sec:kl-def}.

\subsection{Cosmic topology: the cubic three-torus}
\label{sec:topology}

General relativity fixes the local geometry of space but not its global topology, so the Universe may be multiply connected, with a compact fundamental domain whose images tile the covering space~\cite{Lachieze-Rey:1995qrb, COMPACT:2022gbl, COMPACT:2026uzi}. A multiply connected Universe leaves imprints on the cosmic microwave background (CMB), including matched circles in the sky \cite{Cornish:1997ab} and correlations between the CMB temperature and polarization harmonic coefficients that are absent in the covering space \cite{Riazuelo:2003ud,COMPACT:2023rkp,COMPACT:2025adc,MartinBarandiaran:2026qeb}.

The topology experiment compares a cubic three-torus, \E{1}, with its Euclidean covering space, \E{18}. Their local cosmological evolution is the same, but the compact spatial dimensions discretize the allowed Laplacian eigenmodes to wavevectors satisfying
\begin{equation}
\vec k_{\vec n}\cdot\vec T_i=2\pi n_i,\qquad i=1,2,3,
\label{eqn:allowed}
\end{equation}
where the $\vec T_i$ are the translation vectors that generate the topology and $n_i$ are integers. For cubic \E{1}, the three translation vectors are orthogonal and have equal length $L$, making $\theta\equiv L$ the searched parameter. The covering-space integral over Fourier modes therefore becomes a sum over a discrete lattice of $\vec k_{\vec n}$. Because this lattice is not invariant under arbitrary rotations, statistical isotropy is broken, and $\CP(L)$ acquires off-diagonal harmonic correlations even for an isotropic primordial spectrum \cite{Riazuelo:2003ud,COMPACT:2023rkp}. Here we use temperature coefficients with $2\leq\ell\leq40$, so the two hypotheses are
\begin{equation}
Q:\vec a\sim\mathcal N(\vec0,\CQ),\qquad
P(L):\vec a\sim\mathcal N(\vec0,\CP(L)),\qquad
(\CQ)_{\ell m,\ell'm'}=C_\ell^{\LCDM}\delta_{\ell\ell'}\delta_{mm'}.
\label{eq:top_twohyp}
\end{equation}
We hold the cosmological parameters at the Planck 2018 best fit \cite{Planck:2018vyg} under both hypotheses.

We numerically obtain the eigenvalues $\lambda_i(L)$ of $\CQ^{-1/2}\CP(L)\CQ^{-1/2}$ and calculate
\begin{equation}
\DKL(L) =\frac12\sum_i[\lambda_i(L)-1-\ln\lambda_i(L)].
\label{eq:topo-kl}
\end{equation}
The same covariance matrices enter the zero-mean \emph{LLR} of \cref{sec:llr-dist}, which we evaluate for each fixed realization over the whole $L$ grid.

We scan $L\in(\LSS,1.4\,\LSS]$, where $\LSS$ is the diameter of the last-scattering surface. For a cubic torus, $L$ is the shortest separation between the observer and a topological clone, which is an image of the same physical object generated by a nontrivial identification of the covering space. For $L<\LSS$, the last-scattering surface intersects its own topological clone, producing pairs of matched circles \cite{Cornish:1997ab} in the CMB; our lower endpoint excludes this regime. It is consistent with circle-search and likelihood constraints, which require $L>0.985\,\LSS$ \cite{Planck:2013okc,Planck:2015gmu,COMPACT:2022nsu}. As $L$ increases, the eigenmode lattice becomes denser and $\CP(L)$ approaches $\CQ$, with the KL divergence tending to zero as $L\to\infty$.
We fix the orientation of the fundamental domain and search only over its side length $L$; a full search would also scan the three orientation angles \cite{COMPACT:2022gbl}, enlarging the look-elsewhere effect relative to the one-parameter scan used here. The calculation uses full-sky temperature information without instrumental noise or foregrounds.

\subsection{The tensor-to-scalar ratio from B-mode polarization}
\label{sec:rratio}
Primordial tensor perturbations source a curl-like ($B$-mode) component of CMB polarization that scalar perturbations do not produce at linear order~\cite{Kamionkowski:1996zd,Seljak:1996gy}, while gravitational lensing converts $E$ modes into a scalar-sourced $B$-mode foreground~\cite{Zaldarriaga:1998ar}. The polarization experiment compares a model with a primordial tensor contribution to a null containing only lensing $B$ modes and observational noise. We fix the scalar cosmological parameters at the Planck 2018 TT,TE,EE+lowE+lensing best fit~\cite{Planck:2018vyg} and vary the tensor-to-scalar ratio, $\theta\equiv r$. Within the adopted Gaussian approximation, the harmonic coefficients have zero mean and diagonal covariance under both hypotheses, with
\begin{equation}
C_\ell^Q=\mathcal L_\ell+N_\ell,\qquad
C_\ell^P(r)=r\mathcal T_\ell+\mathcal L_\ell+N_\ell,\qquad
\lambda_\ell(r)=1+rs_\ell,\quad s_\ell=\frac{\mathcal T_\ell}{\mathcal L_\ell+N_\ell}.
\label{eq:r-hyp}
\end{equation}
Here $\mathcal L_\ell$ is the lensed-scalar spectrum, $\mathcal T_\ell$ is the tensor spectrum at $r=1$, and $N_\ell$ is the noise spectrum. We obtain the signal spectra from CAMB~\cite{Lewis:1999bs}, fixing $n_t=0$ so that changing $r$ changes only the tensor amplitude. White polarization noise with depth $w_P^{-1}$ and a Gaussian beam is included through~\cite{Knox:1995dq}
\begin{equation}
N_\ell=\left(w_P^{-1}\frac{\pi}{180\times60}\right)^2
\exp\!\left[\frac{\ell(\ell+1)\theta_{\rm FWHM}^2}{8\ln2}\right],
\label{eq:r-noise}
\end{equation}
where $w_P^{-1}$ is in $\mu$K-arcmin and $\theta_{\rm FWHM}$ is in radians.

We consider a LiteBIRD-motivated configuration~\cite{LiteBIRD:2022cnt} with $2\leq\ell\leq40$, full sky ($f_{\rm sky}=1$), and a beam of $\theta_{\rm FWHM}= 30$ arcmin. 
Its effective noise level is chosen to give a Fisher uncertainty~\cite{Tegmark:1996bz} of $\sigma(r)=0.001$ at $r=0$, representing the adopted total uncertainty budget rather than instrumental noise alone. We scan $r\in[0,0.036]$, where the upper end is the current 95\% upper limit from BICEP/Keck combined with Planck and WMAP~\cite{BICEP:2021xfz}, and the exact null, $r=0$, is included in the scan.

Each realization is kept fixed throughout its scan. The \emph{LLR} follows the covariance-shift expression in \cref{sec:llr-dist}.
We treat lensing $B$ modes as Gaussian and apply no delensing. The effective full-sky approximation omits mask-induced correlations and $E$-to-$B$ leakage, while foreground components are not included in our model.

\subsection{Dark energy from supernova distance moduli}
\label{sec:sn}
Type Ia supernova distance moduli provided the first direct evidence for the accelerated expansion of the Universe \cite{Riess:1998cb,Perlmutter:1998np}. 
The supernova experiment compares a cosmological constant with a constant, non-phantom dark energy equation of state. 
For simplicity, we assume a flat cosmology, fix $\Omega_m$ and $\Omega_{\rm DE}=1-\Omega_m$, and
 vary only the dark energy equation-of-state parameter $\theta\equiv w_0$, which is assumed to be constant with redshift.
Here $\Omega_{\rm DE}$ denotes the present dark energy density fraction, including when $w_0\ne-1$. The predicted distance moduli follow from
\begin{align}
E^2(z;w_0)&=\Omega_m(1+z)^3+\Omega_{\rm DE}(1+z)^{3(1+w_0)},\nonumber\\
D_L(z;w_0)&=\frac{c(1+z)}{H_0}\int_0^z\frac{\mathrm dz'}{E(z';w_0)},\qquad
\mu(z;w_0)=5\log_{10}[D_L/\mathrm{Mpc}]+25.
\label{eq:sn-model}
\end{align}
We adopt $\Omega_m=0.315$ and $H_0=67.4\,\mathrm{km\,s^{-1}\,Mpc^{-1}}$, close to the Planck 2018 values~\cite{Planck:2018vyg}, and generate $N=200$ distance moduli at redshifts uniformly spaced over $0.02\leq z\leq1.5$ according to $d_i=\mu(z_i;w_0)+M+n_i$, where $M$ is a common magnitude offset and $n_i\sim\mathcal N(0,\sigma_\mu^2)$, with $\sigma_\mu=0.15$\,mag, are independent errors. The null uses $w_0=-1$,  while alternatives with $w_0>-1$ change the mean distance moduli at the same covariance $\mat C=\sigma_\mu^2\identity$.

We marginalize the common offset $M$ analytically with the same flat prior under both hypotheses \cite{Goliath:2001af,Conley:2011ku}. This also removes the uniform magnitude shift associated with $H_0$, leaving the comparison sensitive to changes in the shape of the distance--redshift relation. Defining $\dmu_w=\vec\mu(w_0)-\vec\mu(-1)$ and $\vec{1}=(1,\ldots,1)^\top$, we calculate
\begin{equation}
\tilde{\mat C}^{-1}=\mat C^{-1}-\frac{\mat C^{-1}\vec{1}\vec{1}^\top\mat C^{-1}}{\vec{1}^\top\mat C^{-1}\vec{1}},\qquad
\DKL(w_0)=\frac12\dmu_w^\top\tilde{\mat C}^{-1}\dmu_w.
\label{eq:sn-Ctilde}
\end{equation}
For our equal, independent magnitude errors, this projection subtracts the average of the components of $\dmu_w$ before computing their squared mismatch, so a constant difference between the models contributes nothing to $\DKL(w_0)$. Equivalently, the comparison uses $N-1$ linearly independent magnitude differences, such as $d_i-d_N$ for $i=1,\ldots,N-1$, in which $M$ cancels. Their joint Gaussian distributions under the two hypotheses are normalized and have KL divergence $\DKL(w_0)$. The \emph{LLR} follows the mean-shift expression in \cref{sec:llr-dist} and is exactly Gaussian at fixed $w_0$, with distributions $\mathcal N(\DKL,2\DKL)$ under the matching alternative and $\mathcal N(-\DKL,2\DKL)$ under $Q$.

We scan $w_0\in[-1,-1/(3\Omega_{\rm DE})]\simeq[-1,-0.487]$ on a grid of 60 points, from the null model with a cosmological constant, $w_0=-1$, to the boundary where the deceleration parameter $q_0=\tfrac12(1+3w_0\Omega_{\rm DE})$ vanishes and present-day acceleration ends. Over this range $\DKL(w_0)$ increases monotonically from zero to $\simeq24.5$. For each alternative model, we draw $2\times10^4$ realizations, and each simulated dataset is kept fixed throughout the scan. We use supernovae alone, with fixed redshifts and independent magnitude errors, leaving correlated systematics and additional BAO or CMB information out of the picture.


\subsection{Calibrating prominence with the false-positive rate}
\label{sec:fpr-calibration}

As discussed above, a null realization can produce a maximum in \emph{LLR} that satisfies the same prominence requirement used to identify a signal. We therefore apply the search procedure of \cref{sec:peak} to realizations drawn from $Q$ and measure the fraction that contain an accepted peak. Following \cref{eq:prominence,eq:detection-criterion}, the false-positive rate ($\FPR$) and the detection probability, or true-positive rate ($\TPR$), for experiment $e$ are
\begin{equation}
\begin{aligned}
\FPR_e(\mathrm{prom}_{\mathrm{thr}})
&=\Pr_Q\!\left[\mathrm{prom}(\hat\theta)\geq\mathrm{prom}_{\mathrm{thr}}\right],\\
\TPR_e\left(\DKL(\theta_e),\mathrm{prom}_{\mathrm{thr}}\right)
&=\Pr_{P(\theta_e)}\!\left[\mathrm{prom}(\hat\theta)\geq\mathrm{prom}_{\mathrm{thr}}\right],
\end{aligned}
\label{eq:experiment-rates}
\end{equation}
where $\hat\theta$ is the location of the highest eligible peak of the profile, as defined in \cref{eq:detection-criterion}, and $\mathrm{prom}(\hat\theta)$ is its prominence.
Here, $\theta_e$ denotes the underlying parameter from which the data are generated for experiment $e$. Since $\theta$ and the KL divergence $\DKL(\theta_e)$ are monotonically related, we use $\DKL(\theta_e)$ as the argument of the $\TPR$ to provide a common measure of signal strength across experiments.
Both rates are calculated over all realizations of the relevant model, including those for which no eligible peak is found. With $N_Q$ null realizations, the Monte Carlo uncertainty on a rate $p$ is $\sqrt{p(1-p)/N_Q}$, so rates well below $1/N_Q$ cannot be resolved.

\begin{figure}[t]
\centering
\includegraphics[width=0.82\linewidth]{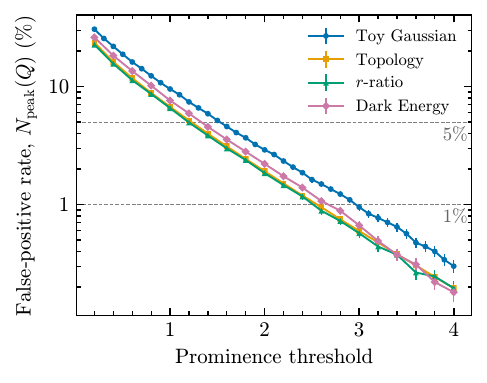}
\caption{False-positive rate as a function of the prominence threshold for the controlled Gaussian model (blue), cubic three-torus (orange), $B$-mode experiment (green), and supernova experiment (pink). Each rate is the percentage of null realizations for which the search accepts an eligible peak. The vertical axis is logarithmic. Horizontal dashed lines mark the $\sim 5\%$ and $\sim 1\%$ reference rates.}
\label{fig:fpr-vs-threshold}
\end{figure}

\Cref{fig:fpr-vs-threshold} shows that the false-positive rate decreases smoothly with the prominence threshold in all four experiments. Over much of the plotted range, the curves are approximately straight on the logarithmic vertical scale, indicating an approximately exponential decrease. A similar exponential tail is expected for the height of the maximum of a smooth random process, whose exceedance probability at high levels is dominated by an up-crossing term decaying as $e^{-u}$ in units of the \emph{LLR} \cite{Davies:1987zz,Gross:2010qma}; since our statistic is the prominence rather than the height, we use this only as a qualitative guide. Increasing $\mathrm{prom}_{\mathrm{thr}}$ therefore provides a practical way to reduce false detections, while the null ensemble allows us to determine the threshold corresponding to a desired false-positive rate.

The curves are similar but do not coincide. From the plotted crossings, a $5\%$ false-positive rate requires a threshold of approximately $1.5$ for the toy Gaussian model and $1.2$ for the other three experiments. Reducing the target to $1\%$ raises these values to about $3$ and $2.5$, respectively. These differences show why prominence must be calibrated for each searched family of alternatives if the experiments are to be compared at a common false-positive rate.

For the peak-location comparison below, we therefore choose the thresholds experiment by experiment so that they correspond to the same false-positive rates. The $5\%$ cut, a relatively permissive criterion, is $\mathrm{prom}_{\mathrm{thr}}=1.2$ for the cubic three-torus, $r$-ratio, and constant dark energy experiments and $1.5$ for the toy Gaussian model. The $1\%$ cut, a stricter criterion, is $\mathrm{prom}_{\mathrm{thr}}=2.5$ and $3$, respectively. We also consider a third, common threshold, $\mathrm{prom}^{\star}_{\mathrm{thr}}=0.2$, which places the low-$\DKL({\rm true})$ plateau of the mean accepted peak divergence close to unity in all four experiments. This third cut addresses a different question: how often the search accepts weak-signal realizations when their mean selected peak location is near $\DKL({\rm peak})=1$. Because the false-positive rate decreases with increasing threshold, this cut necessarily has a higher false-positive rate around $20\%$--$30\%$.

\subsection{Peak locations and detection probabilities}
\label{sec:peak-calibration}

\begin{figure}[t]
\centering
\includegraphics[width=\textwidth]{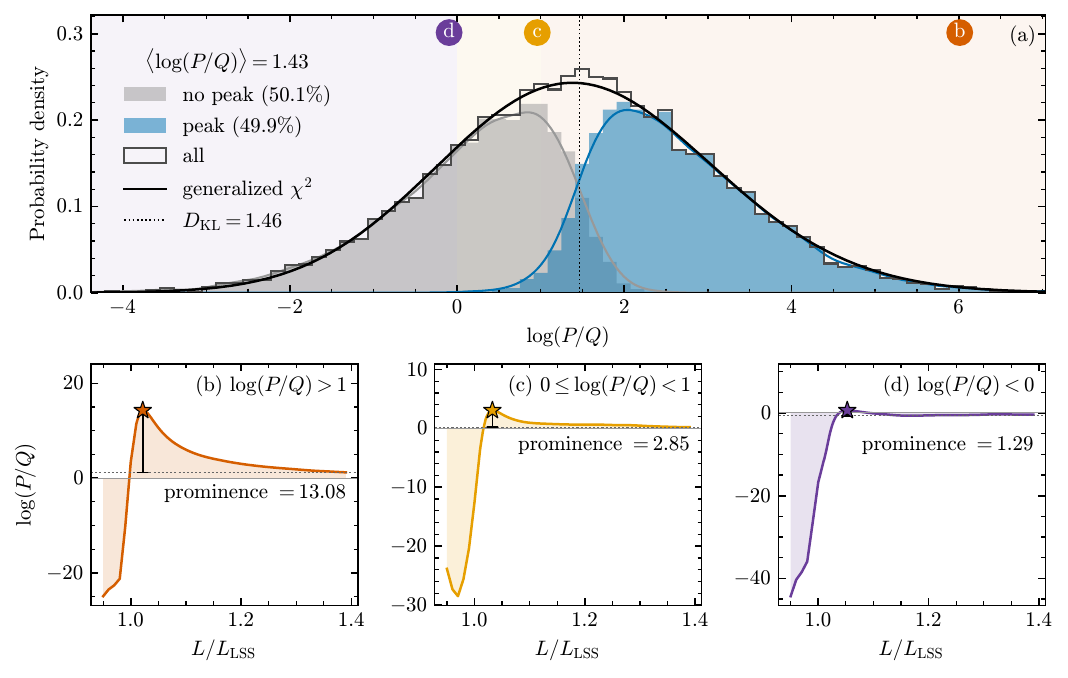}
\caption{The $L=1.1\,\LSS$ example of cubic \E{1} topology with $\KL{\E{1}}{\E{18}}= 1.46$.
(a)~Distribution of $\log(P/Q)$ over $n_{\rm real}=20{,}000$ realizations
drawn under $P$, split into realizations with and without a detected peak, together with the single generalized-$\chi^2$ fit (black) and separate kernel density estimates for each subpopulation.
(b)--(d)~Individual profile scans of $\LR(\vec{a};\theta)$ against
$L/\LSS$ for realizations drawn from the $\log(P/Q)>1$,
$0<\log(P/Q)<1$, and $\log(P/Q)<0$ subsets, respectively. 
In panel (a), the corresponding ranges of $\log(P/Q)$ are labeled and colored to match the appropriate panel (b)-(d). 
} 
\label{fig:example-profiles}
\end{figure}

Having established how the prominence cuts affect false positives, we next examine which parameter value is inferred when the data are generated from a specific alternative model. 
\Cref{fig:example-profiles} illustrates a representative case for non-trivial topology with $L=1.1\LSS$, corresponding to $\KL{P}{Q}=1.46$, where $P$ is a universe with cubic topology and $Q$ is the standard covering space, with it's trivial topology. Panel~(a) shows the distribution of $\log(P/Q)$ for the $n_{\rm real}=20{,}000$ realizations drawn from $P$, separated according to whether the realizations have detected a peak using a prominence threshold corresponding to $5\%$ false positive rate. With this criterion, around half of the realizations contain a detected peak. The sample mean of $\log(P/Q)$, corresponding to an estimated $\KL{P}{Q}\approx 1.43$, agrees with the input value of $1.46$ within the expected Monte Carlo uncertainty, confirming that the simulated realizations provide an accurate sampling of $\CP(\theta_{\rm true})$.
Panels~(b)--(d) show representative likelihood profiles, $\LR(\vec{a};\theta)$ as a function of $L/\LSS$, drawn from realizations with $\log(P/Q)>1$, $0<\log(P/Q)<1$, and $\log(P/Q)<0$, respectively. In each panel, the peak is marked with a star, and its prominence is listed in the legend ($13.08$, $2.85$, and $1.29$, respectively).
Panel~(d) of \cref{fig:example-profiles} shows that even a realization with $\log(P/Q)<0$, still can produce a prominent peak.

\begin{figure*}[t]
\centering
\includegraphics[width=\textwidth]{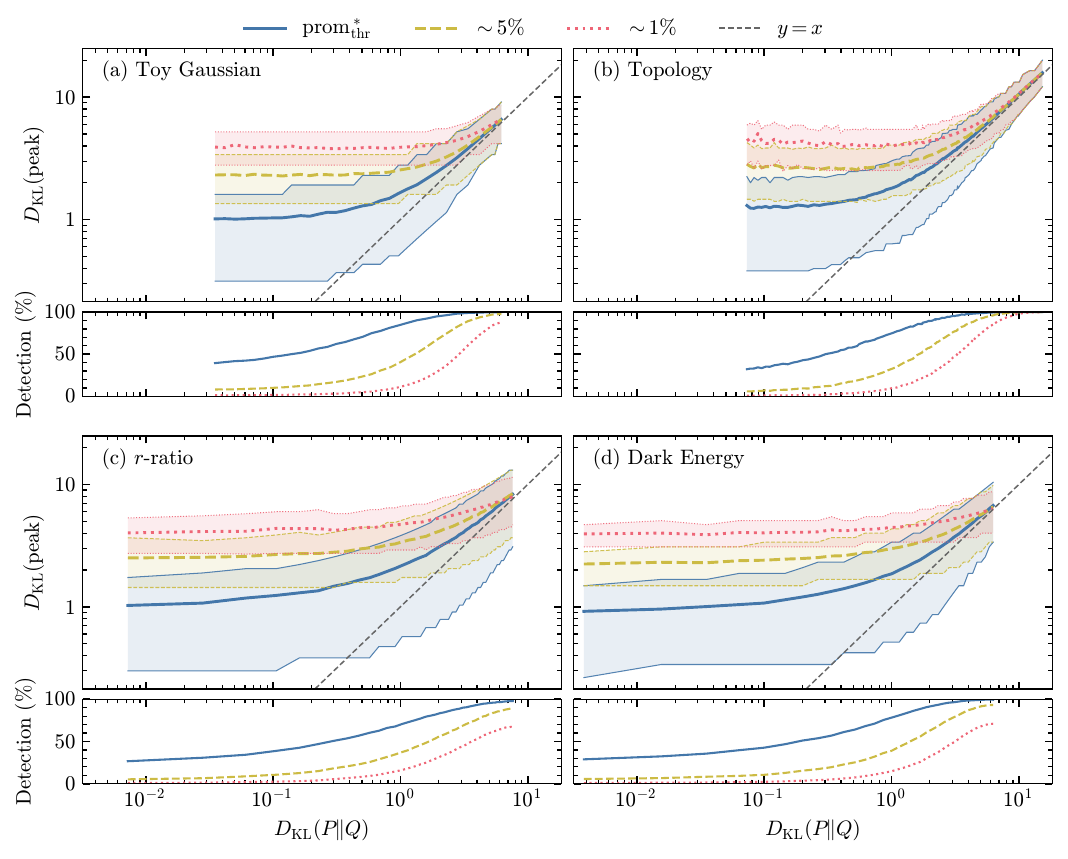}
\caption{Locations of accepted likelihood-profile peaks and their detection fractions for (a) the toy Gaussian model, (b) cubic three-torus topology, (c) the tensor-to-scalar ratio from $B$ modes, and (d) dark energy from supernovae. The curves in the upper panels show the KL divergence of the selected models, $\DKL({\rm peak})$, against the generating KL divergence, $\DKL({\rm true})=\KL{P(\theta_{\rm true})}{Q}$, with both axes logarithmic. The shaded regions span the 16th--84th percentiles of $\DKL({\rm peak})$, showing the spread in the KL divergences of the selected models rather than the uncertainty on their mean. Both the means and percentiles include only realizations that pass the peak-selection criterion. Purple solid, blue dashed, and orange dotted curves correspond to $\mathrm{prom}^{\star}_{\mathrm{thr}}=0.2$ and to the thresholds giving $5\%$ and $1\%$ false-positive rates in each experiment; 
the gray dashed diagonal marks $\DKL({\rm peak})=\DKL({\rm true})$. The lower panels show the percentage of all realizations accepted at each threshold.}
\label{fig:peak-kl-calibration}
\end{figure*}

So far we have only asked \emph{whether} a peak is found; we have not asked \emph{where in parameter space} it lands relative to the alternative model that actually generated the data. In \cref{fig:peak-kl-calibration}, the curves in the upper panels show the mean KL divergence of the models selected by the peak locations, $\langle \DKL({\rm peak})\rangle$, against the KL divergence of the underlying model that generates the data, $\DKL({\rm true})$. The shaded regions in the upper panels span the 16th--84th percentiles of $\DKL({\rm peak})$ among accepted realizations at each $\DKL({\rm true})$, showing the spread of peak locations rather than the uncertainty on their mean. The diagonal dashed line marks equality between these quantities, while the lower panels show the fraction of realizations with a detected peak at each prominence threshold. The two rows should be read together, since the peak-location averages include only the realizations accepted by the search.

At the $\sim5\%$ and $\sim1\%$ false-positive thresholds in \cref{fig:peak-kl-calibration}, all four experiments show similar behavior when $\DKL({\rm true})$ is small. As $\DKL({\rm true})$ decreases below roughly $5$--$7$, the mean inferred KL divergence, $\langle \DKL({\rm peak})\rangle$, departs from the diagonal and gradually approaches a plateau. For the $5\%$ cut, the plateau lies around $2$--$3$, rising to about $4$ for the $1\%$ cut. The corresponding true-positive rates remain small in the low-$\DKL({\rm true})$ regime.

\begin{figure}
    \centering
    \includegraphics[width=0.9\columnwidth]{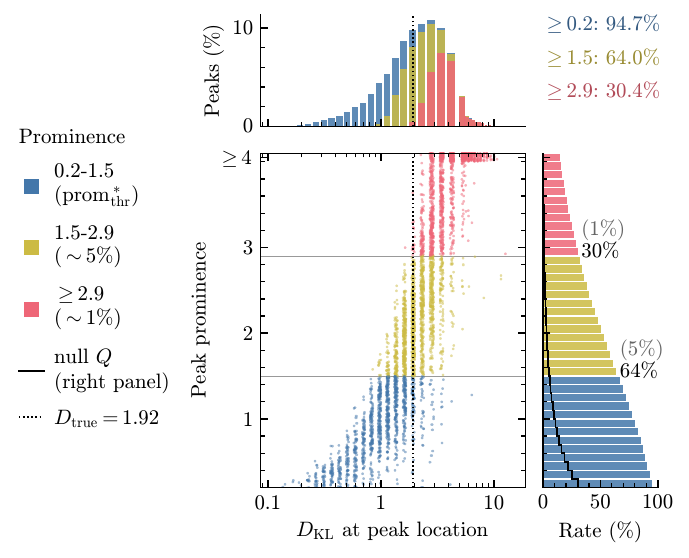}
    \caption{Peak prominence versus peak location for the toy Gaussian model with $\DKL({\rm true})=1.92$ ($N=20{,}000$ realizations; $4000$ points shown). The gray lines mark prominence thresholds corresponding to false-positive rates of approximately $5\%$ and $1\%$; the loosest threshold, $\mathrm{prom}^\star_{\mathrm{thr}}=0.2$, defines the lower edge of the panel. Each point represents one realization, colored by the strictest threshold its peak passes: blue peaks pass only the loosest threshold, yellow peaks also pass the $5\%$ threshold, and red peaks pass the $1\%$ threshold. The dotted line marks $\DKL({\rm true})$. \textbf{Top:} Nested histograms of accepted peak locations, expressed as percentages of all realizations, with total acceptance rates at upper right. \textbf{Right:} Fractions of realizations with peaks at or above each prominence under the alternative (bars) and null (black outline). Labels give the true-positive rates, with false-positive rates in parentheses.}
    \label{fig:peak_joint}
\end{figure}
 
The common threshold $\mathrm{prom}^{\star}_{\mathrm{thr}}=0.2$ places this plateau close to unity in all four experiments. These comparisons show that realizations of models with small $\DKL({\rm true})$ can still exhibit an accepted peak within the explored parameter domain. However, at the $5\%$ and $1\%$ $\FPR$ thresholds, their mean peak locations correspond to substantially larger KL divergences than those of the generating models. Thus, the presence of a detected peak does not, by itself, imply that the underlying model lies near the inferred location in parameter space. This is a selection effect analogous to Eddington bias: conditioning on a threshold preferentially retains realizations whose fluctuations pushed the peak upward \cite{Eddington:1913}, and intervals whose construction depends on whether a detection was claimed can lose their nominal coverage \cite{Feldman:1997qc}.

Comparing the thresholds in \cref{fig:peak-kl-calibration} shows that a stricter prominence requirement raises the plateau of the mean accepted peak divergence in the low-$\DKL({\rm true})$ regime while reducing the true-positive rate. Consequently, the realizations that survive the stricter criterion favor models with larger $\DKL({\rm peak})$. The location of the plateau therefore depends on the selection criterion, as well as on the family of models being searched.

As $\DKL({\rm true})$ increases, the mean $\DKL({\rm peak})$ approaches the dashed diagonal, indicating closer agreement, on average, between the KL divergences of the inferred and generating models. The true-positive rates also increase, approaching unity within the displayed range for the toy Gaussian and topology experiments. Nevertheless, agreement between the mean divergences does not guarantee accurate parameter recovery for an individual realization. 

To examine this more directly, \cref{fig:peak_joint} shows the joint distribution of peak prominence and $\DKL({\rm peak})$ for the toy Gaussian model with $\DKL({\rm true})=1.92$. Although all realizations come from the same model, their $\DKL({\rm peak})$ values span a broad range for a given prominence. In this example, the two quantities are strongly related; in other words, larger prominence biases the result toward larger $\DKL({\rm peak})$. In addition to the loosest threshold of $0.2$, we consider prominence thresholds of $1.5$ and $2.9$, which give false-positive rates of approximately $5\%$ and $1\%$, respectively. Each point represents one realization ($4000$ shown), colored by the strictest threshold its peak passes: blue peaks pass only the loosest threshold, yellow peaks also pass the $5\%$ threshold, and red peaks pass the $1\%$ threshold.

The upper panel shows the distributions of accepted peak locations, normalized to all realizations. The histograms are nested, so the visible blue and yellow portions represent peaks excluded by the $5\%$ and $1\%$ thresholds, respectively. Approximately $95\%$, $64\%$, and $30\%$ of realizations pass the three thresholds, respectively. As the threshold increases, fewer peaks are accepted, and their locations become more concentrated above $\DKL({\rm true})$. The right panel shows the fractions of alternative and null realizations with peaks at or above each prominence threshold, giving the true-positive and false-positive rates.

Consequently, a peak at, say, $\DKL({\rm peak})\sim4$ can arise from a broad range of generating models, including models with $\DKL({\rm true})\lesssim0.1$. The peak location alone is therefore insufficient to determine which of these models generated the data. Any uncertainty estimate derived from that location must account for its distribution across a wider range of alternatives, including low-$\DKL({\rm true})$ models that are difficult to distinguish from the null. If additional features of the likelihood profile are used to narrow this range, their ability to distinguish the generating models must likewise be calibrated against simulated realizations.

\subsection{ROC comparison across the four experiments}
\label{sec:experiment-roc}

\begin{figure*}[t]
\centering
\includegraphics[width=\textwidth]{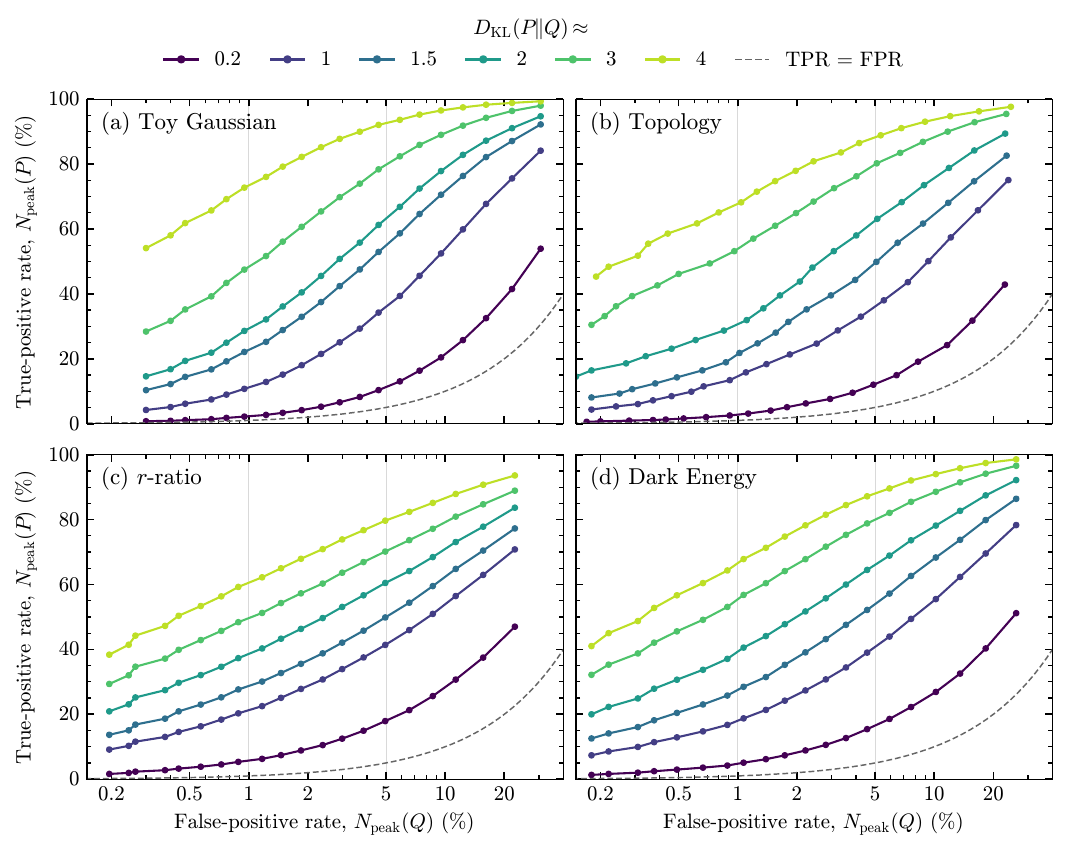}
\caption{ROC curves for (a) the toy Gaussian model, (b) cubic three-torus topology, (c) the tensor-to-scalar ratio from $B$ modes, and (d) dark energy from supernovae. Each curve shows the true-positive rate (TPR) against the false-positive rate (FPR) for a fixed generating model as the prominence threshold is varied. Colors distinguish the generating KL divergences, $\DKL({\rm true})\approx0.2$, 1, 1.5, 2, 3, and 4 in every panel, and markers indicate the evaluated thresholds. Both rates are expressed as percentages, with a logarithmic horizontal axis and a linear vertical axis. The gray dashed curve marks equal true- and false-positive rates.}
\label{fig:roc-curves}
\end{figure*}

Having examined which parameter value is inferred under our detection criterion, we next compare how often the search detects a peak under the alternative and under the null, or equivalently, how $\TPR$ varies with $\FPR$ for different values of $\DKL({\rm true})$. In \cref{fig:roc-curves}, the ROC curves show the true-positive rate against the false-positive rate as the prominence threshold is varied. As shown in \cref{fig:fpr-vs-threshold}, increasing this threshold reduces the false-positive rate. Each curve corresponds to a fixed generating model, with its KL divergence, $\DKL({\rm true})$, given in the legend. Moving along a curve therefore changes the prominence threshold, while moving between curves changes the model that generates the data.

All four experiments show the same overall trend: at a fixed false-positive rate, the true-positive rate increases with $\DKL({\rm true})$. For a given generating model, increasing the prominence threshold reduces both rates. Thus, suppressing false detections also excludes some realizations generated by the alternative, and the ROC curves show how this trade-off depends on the information available in the data.

For models with $\DKL({\rm true})\simeq1$, a $1\%$ false-positive rate gives a true-positive rate of only about $10$--$20\%$ across the four experiments. Allowing a $5\%$ false-positive rate raises this to roughly $30$--$40\%$, but a substantial fraction of alternative realizations still contain no accepted peak. Consequently, reaching a KL divergence of unity does not imply that a typical realization will produce an accepted peak. Detection depends on the prominence criterion and the corresponding false-positive rate used in the search.

At larger generating KL divergences, the alternative models are detected more frequently. For $\DKL({\rm true})\simeq4$, the true-positive rates are approximately $60$--$75\%$ at a $1\%$ false-positive rate, rising to about $80$--$95\%$ at a $5\%$ false-positive rate. Although the curves show similar behavior across all four experiments, they do not coincide at comparable generating KL divergences.

These results complement the peak-location comparison. For models with small $\DKL({\rm true})$, an accepted peak can favor a model with substantially larger KL divergence, while the ROC curves show how infrequently such detections occur at a chosen false-positive rate. Moreover, even when the generating KL divergence is near unity, most realizations can remain undetected under a stringent prominence threshold. The interpretation of a detected peak therefore requires both its location and the calibrated detection and false-positive rates; neither the presence of the peak nor the KL divergence associated with it is sufficient on its own.

\section{Conclusions}
\label{sec:conclusions}

The KL divergence measures the expected log-likelihood ratio between two probability distributions (hypotheses), providing a non-negative average measure of their distinguishability. However, interpreting it as the discovery potential of an alternative model requires a specified detection criterion and its associated true-positive and false-positive rates. For Gaussian distributions, we derived the log-likelihood-ratio (\emph{LLR}) distribution and its cumulants under both hypotheses. When the models differ only in their means, the \emph{LLR} is exactly Gaussian, and its distribution is fully determined by the KL divergence. When the model information lies in their covariances, the \emph{LLR} instead follows a generalized $\chi^2$ distribution, whose width and higher cumulants contain additional information that is not captured by the KL divergence alone.

In this work, we studied four families of models: a toy Gaussian model, cubic three-torus cosmic topology, the tensor-to-scalar ratio inferred from $B$-mode polarization, and dark energy inferred from supernova distances. For each family, we searched over one unknown parameter and used the prominence of the highest likelihood-profile peak as an operational detection statistic, measuring how far the peak rises above its surrounding baseline. In the four examples considered in this paper, the null model lies at a boundary or limiting value of the parameter space. Such situations arise commonly when a model is extended by adding non-negative parameters whose vanishing values recover the null, and require calibration beyond the usual interior-point assumptions of Wilks' theorem. Searching over many parameter values gives random fluctuations in the data more opportunities to produce an eligible peak, so the peak's statistical significance must be assessed using the same complete search on simulated null data. We addressed both issues by calibrating the prominence threshold against null simulations using the same scanned range and grid, thereby accounting for boundary effects and the look-elsewhere effect without assuming an asymptotic $\chi^2$ distribution for the detection statistic or estimating a separate trials factor.

Across these experiments, when we apply the search to simulated data generated by alternative models with $\DKL({\rm true})\simeq 1$ with respect to the null, the detection probabilities (true-positive rates) are only about $10$--$20\%$ at a $1\%$ false-positive rate, rising to approximately $60$--$75\%$ for models with $\DKL({\rm true})\simeq 4$. Allowing a $5\%$ false-positive rate increases these probabilities to roughly $30$--$40\%$ and $80$--$95\%$, respectively. Therefore, a KL divergence of unity does not imply that a typical realization will contain an accepted peak. Interestingly, all four experiments show these trends, although their detection probabilities do not coincide at comparable KL divergences.

For parameter estimation, the more consequential result is that the model selected by an accepted peak can have a substantially larger KL divergence than the model that generated the data. Although alternatives with $\DKL({\rm true})\lesssim2$ can produce prominent peaks, the parameter values inferred from those accepted peaks tend to lie in regions farther from the null. Quantitatively, in our experiments, accepted peaks tend to occur at $\DKL({\rm peak})\gtrsim1$, with the preferred region shifting to $\DKL({\rm peak})\gtrsim2$ at a $5\%$ false-positive rate and roughly $\DKL({\rm peak})\gtrsim3$ at a $1\%$ rate. Thus, when KL divergence is used as a practical guideline for where a search may yield an accepted parameter estimate, our analysis suggests a characteristic scale of a few nats, depending on the required false-positive rate. These values describe approximate regions of accepted peak locations, rather than sharp lower bounds or thresholds for accurate parameter recovery. Stricter prominence requirements amplify the selection effect and shift the accepted estimates toward larger divergences.

The distinction between $\DKL({\rm true})$ and $\DKL({\rm peak})$ is therefore essential when using this guideline. An inferred parameter with a KL divergence of a few nats does not, by itself, exclude models with much smaller KL divergences, since data drawn from those models can also produce an accepted peak at that location. Consequently, a KL-based guideline must not be used to restrict uncertainty estimates to the large-divergence region in which detections preferentially occur. Assessing frequentist coverage requires accounting for the distribution of inferred parameters under weakly distinguishable alternatives, including any conditioning on detection, rather than relying only on the local shape of the accepted peak. In a Bayesian analysis, credible intervals likewise require evaluating the posterior over the relevant parameter domain under the adopted prior, including low-divergence regions rather than excluding them on the basis of a KL divergence threshold.

We have developed a prescription for using KL divergence and log-likelihood-ratio distributions to forecast discovery potential and assess parameter recovery in a continuous parameter search. By combining established statistical tools, this prescription clarifies what a given KL divergence implies for the detection probability and how reliably accepted parameter estimates recover the underlying model that generated the data. Our simple cosmological case studies illustrate how these outcomes depend on the model family, parameter boundaries, and detection criterion.

\section*{Acknowledgments}
This work made use of the High-Performance Computing Resource in the Core Facility for Advanced Research Computing at Case Western Reserve University and the facilities of the Ohio Supercomputing Center.
D.P.M., G.D.S., and A.K.\ acknowledge support from NASA ATP grant 80NSSC22K0805.
D.P.M and G.D.S. acknowledge support from NASA ADAP Grant 24-ADAP24-0018. 
A.S.\ and D.P.M.\ acknowledge support by the Bulgarian National Science Fund programme ``VIHREN--2024'' project No. KP--06--DV/9/17.12.2024.
M.M.B.\ acknowledges support by the Spanish Ministry of Science, Innovation and Universities under the FPU predoctoral grant \\ FPU22/02306.
J.C.D.\ is supported by the Spanish Research Agency (Agencia Estatal de Investigaci\'on), the Ministerio de Ciencia, Innovaci\'on y Universidades, and the European Social Funds through grant JDC2023-052152-I, as part of the Juan de la Cierva programme.
G.D.S.\ and A.S.\ acknowledge past support from DOE grant DESC0009946.
Y.A.\ acknowledges support by the Spanish Research Agency (Agencia Estatal de Investigaci\'on)'s grant RYC2020-030193-I/AEI/10.13039/501100011033, by the European Social Fund (Fondo Social Europeo) through the Ram\'{o}n y Cajal programme within the State Plan for Scientific and Technical Research and Innovation (Plan Estatal de Investigaci\'on Cient\'ifica y T\'ecnica y de Innovaci\'on) 2017-2020, by the Spanish Research Agency through the grant IFT Centro de Excelencia Severo Ochoa No CEX2020-001007-S funded by MCIN/AEI/10.13039/501100011033, by the Spanish National Research Council (CSIC) through the Talent Attraction grant 20225AT025, and by the Spanish Research Agency's Consolidaci\'on Investigadora 2024 grant CNS2024-154430.
F.C.G. is supported by Ministerio de Ciencia, Innovaci\'on y Universidades, Spain, through a Beatriz Galindo Junior grant BG23/00061 and by an UCOLIDERA grant from the Universidad de C\'ordoba.
A.H.J.\ acknowledges support from STFC in the UK.
A.N.\ and C.T.\ are supported by Richard S.\ Morrison Fellowships.
A.T.\ is supported by the European Union's Horizon Europe research and innovation
programme under the Marie Sk\l{}odowska-Curie grant agreement No.\ 101126636.
M.M.B. thanks CWRU for the hospitality while working on this project.

\bibliographystyle{utphys}
\bibliography{topology, additional}

\end{document}